\documentclass[sigconf]{acmart}

\usepackage{physics}
\usepackage{algorithm}
\usepackage[noend]{algpseudocode}
\usepackage{graphicx}
\usepackage{subcaption}

\begin{document}
\title{Modeling Multidimensional User Relevance in IR using Vector Spaces}

\copyrightyear{2018} 
\acmYear{2018} 
\setcopyright{othergov}
\acmConference[SIGIR '18]{The 41st International ACM SIGIR Conference on Research \& Development in Information Retrieval}{July 8--12, 2018}{Ann Arbor, MI, USA}
\acmBooktitle{SIGIR '18: The 41st International ACM SIGIR Conference on Research \& Development in Information Retrieval, July 8--12, 2018, Ann Arbor, MI, USA}
\acmPrice{15.00}
\acmDOI{10.1145/3209978.3210130}
\acmISBN{978-1-4503-5657-2/18/07}

\author{Sagar Uprety}
\affiliation{%
  \institution{The Open University}
  \streetaddress{Walton Hall}
  \city{Milton Keynes}
  \country{United Kingdom}
  \postcode{MK7 6AA}
}
\email{sagar.uprety@open.ac.uk}

\author{Yi Su}
\affiliation{%
  \institution{ School of Computer Science and Technology, Tianjin University}
  \streetaddress{P.O. Box 1212}
  \city{Tianjin}
  \country{China}
  \postcode{43017-6221}
}
\email{suyi2016@tju.edu.cn}

\author{Dawei Song}
\affiliation{%
  \institution{The Open University}
  \streetaddress{Walton Hall}
  \city{Milton Keynes}
  \country{United Kingdom}
} 

\additionalaffiliation{%
  \institution{ Beijing Institute of Technology. Correspondence Author}
  \streetaddress{Haidian District}
  \city{Beijing}
  \country{China}
  \postcode{43017-6221}
}
\email{dawei.song@open.ac.uk}

\author{Jingfei Li}
\affiliation{%
  \institution{National Computer Network Emergency Response Technical Team/Coordination Center of China}
  \streetaddress{Chaoyang District}
  \city{Beijing}
  \country{China}
  \postcode{}
}
\additionalaffiliation{%
  \institution{ Tianjin University. Correspondence author}
  \streetaddress{P.O. Box 1212}
  \city{Tianjin}
  \country{China}
  \postcode{43017-6221}
}

\email{jingfl@foxmail.com}

\renewcommand{\shortauthors}{S. Uprety et al.}

\begin{abstract}
It has been shown that relevance judgment of documents is influenced by multiple factors beyond topicality. Some multidimensional user relevance models (MURM) proposed in literature have investigated the impact of different dimensions of relevance on user judgment. Our hypothesis is that a user might give more importance to certain relevance dimensions in a session which might change dynamically as the session progresses. This motivates the need to capture the weights of different relevance dimensions using feedback and build a model to rank documents for subsequent queries according to these weights. We propose a geometric model inspired by the mathematical framework of Quantum theory to capture the user's importance given to each dimension of relevance and test our hypothesis on data from a web search engine and TREC Session track.
\end{abstract}

%
%



\maketitle

\section{Introduction}
There is a growing body of work investigating different factors which affect user's judgment of relevance~\cite{ASI:Barry-relevance, ASI:Tombros-relevance, ASI:Xu-relevance, ASI:Beresi, MURM-psychometrics, ASI:Xu-MURM, ASI:Jingfei,10.1007/978-3-642-00958-7_25}. A multidimensional user relevance model (MURM) was proposed \cite{ASI:Xu-MURM,MURM-psychometrics} which defined five dimensions of relevance namely "Novelty", "Reliability", "Scope", "Topicality" and "Understandability". In a recent paper~\cite{ASI:Jingfei} an extended version of the MURM comprising two additional dimensions "Habit" and "Interest" is proposed. The "Interest" dimension refers to the topical preferences of users in the past, while "Habit" refers to their behavioral preferences. For example, accessing specific websites for some particular information or task is considered under the "Habit" dimension. Experiments on real-world data show that certain dimensions, such as "Reliability" and "Interest", are more important for the user than "Topicality", in judging a document.

Our hypothesis is that in a particular search session or search task, there is a particular relevance dimension or a combination of relevance dimensions which the user has in mind before judging documents. For example, if the user wants to get a visa to a country, he or she would prefer documents which are more reliable ("Reliability") for this task, but when looking to book flights to that country, the user might go to his or her preferred websites ("Habit"). Therefore, for next few queries of the session, "Habit" dimension becomes more important. Thus, the importance given to relevance dimensions might change as the session progresses or tasks switch. By capturing the importance assigned to each dimension for a query, we can model the dimensional importance and use it to improve the ranking for the subsequent queries. The relevance dimensions are modeled using the Hilbert space formalism of Quantum theory which unifies the logical, probabilistic and vector space based approaches to IR \cite{Rijsbergen:2004:GIR:993731}. We place the user's cognitive state with respect to a document at the center of the IR process. Such a state is modeled as an abstract vector with multiple representations existing at the same time in different basis corresponding to different relevance dimensions. This cognitive state comes into reality only when it is measured in the context of user interactions.

\begin{figure*}[t]

   \begin{subfigure}[t]{0.32\textwidth}
        \includegraphics[width=\textwidth]{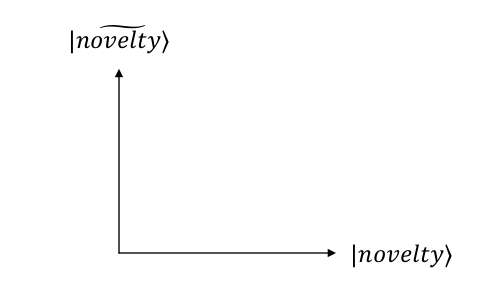}
        \caption{Figure 1.a}
   \end{subfigure}  
    \begin{subfigure}[t]{0.33\textwidth}
        \includegraphics[width=\textwidth]{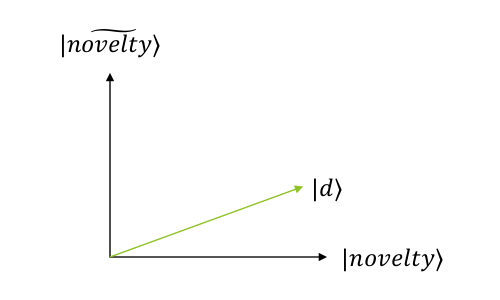}
        \caption{Figure 1.b}
    \end{subfigure}
    \begin{subfigure}[t]{0.34\textwidth}
        \includegraphics[width=\textwidth]{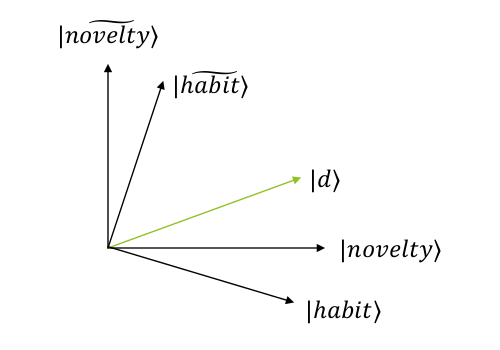}
         \caption{Figure 1.c}
  \end{subfigure}

\end{figure*} 

\section{Geometric representation of multidimensional relevance}
Consider a real-valued two dimensional Hilbert Space. Relevance with respect to a dimension (e.g. Reliability) is a vector in this Hilbert space. Non-relevance with respect to the same dimension is an orthogonal vector to it. Further, we denote vectors as kets following the Dirac's notation. For example, the vectors for relevance and non-relevance with respect to novelty are denoted as $\ket{novelty}$ and $\ket{\widetilde{novelty}}$. Figure 1.a shows the construction of a two-dimensional Hilbert Space for a relevance dimension. 

Next, we model the user's perception of a document with respect to a dimension of relevance also as a vector in this Hilbert space. This vector is a superposition of relevance and non-relevance vectors with respect to a dimension, e.g., $\ket{d} = \alpha\ket{novelty} + \beta\ket{\widetilde{novelty}}$. The coefficient $|\alpha|^2$ is the weight (i.e., probability of relevance) the user assigns to document $d$ in term of novelty, and $|\alpha|^2 + |\beta|^2 = 1$. We will talk about how to calculate these coefficients in the next section. Figure 1.b shows the modeling of user's cognitive state for document $d$ with respect to the Novelty dimension. 

Depending on a user's preference of relevance dimensions for a particular query, the user will judge the same document differently. A document might be of interest to the user but may not be novel when the user is looking for latest documents about the query. This phenomena can be modeled in the same Hilbert space by having different basis for different dimensions of relevance. The same document $d$ can be written in terms of another set of basis vectors corresponding to another dimension of relevance. For example:
\begin{align}
\ket{d_1} &= \alpha_{11}\ket{novelty} + \beta_{11}\ket{\widetilde{novelty}} \nonumber \\
	      &= \alpha_{12}\ket{habit} + \beta_{12}\ket{\widetilde{habit}} \nonumber \\
	      &= \alpha_{13}\ket{topic} + \beta_{13}\ket{\widetilde{topic}}
\end{align}
 and so on in all seven basis. Figure 1.c shows the construction of such a Hilbert space showing two basis for simplicity. 

We have represented user's cognitive state with respect to a single document in different basis corresponding to different dimensions of relevance. Similarly, we can do that for all the documents retrieved for a query. Each document will be represented in a separate Hilbert space.

The user's cognitive state for a document $d$ is an abstract vector, because the vector has different representations in different basis. It does not have a definite state, and a particular representation comes into picture only when we talk of a particular relevance dimension. This is similar to the concept of a state vector in Quantum theory which contains all the information about a quantum system, yet is an abstract entity and has different representations of the same system. We get to see a particular representation of a system depending on how we measure it. A document may look highly relevant in one basis, if it has a high weight in that basis and the user makes judgment from the perspective of that relevance dimension. However, the relevance can change if the user considers a different basis (a different perspective of looking at the document).

\section{Capturing user's changing weights to relevance dimensions}
Having established the geometric representation of documents, we can make use of it to capture the weights of relevance dimensions for a user in response to a query (Algorithm 1). 

In Algorithm 1, the input parameters $docsALL$ and $docsSAT$ correspond to the list of all retrieved documents and SAT-clicked~\cite{ASI:Jingfei} documents for a query respectively. We quantitatively define each of the seven relevance dimensions using some features. For each query-document pair, these features are extracted and computed (Step 3). They are integrated into the LambdaMART~\cite{Burges2010FromRT} Learning to Rank(LTR) algorithm to generate seven relevance scores (one for each dimension) for the query-document pair (Step 4). Please refer to \cite{ASI:Jingfei} for more details about the features defined for each dimension and the hyper-parameters of the LTR algorithm. Thus, for a query and its set of retrieved documents, we have seven different rankings, one for each relevance dimension. The seven scores assigned to a document for a query are then normalized using the min-max normalization technique across all the documents for the query (Step 5). The normalized score for each dimension forms the coefficient of superposition of the relevance vector for the respective dimension. It quantitatively represents user's perception of the document for that dimension. For example, for a query $q$, let $d_1, d_2, ..., d_n$ be the ranking order corresponding to the "Interest" dimension. Let relevance scores be $\lambda_1, \lambda_2, ..., \lambda_n$ respectively. We construct the vector for document $d_1$ in the 'Interest' basis as:
\begin{equation}
\ket{d_1} = \alpha_{11}\ket{interest} +\beta_{11}\ket{\widetilde{interest}}
\end{equation} 
where $\alpha_{11} = \sqrt{\frac{\lambda_1 - min(\lambda)}{max(\lambda) - min(\lambda)}}$, where $max(\lambda)$ is the maximum value among $\lambda_1, \lambda_2, ..., \lambda_n$. Square root is taken in accordance with the quantum probability framework. Similarly, the second document is represented in its Hilbert space as:
\begin{equation}
\ket{d_2} = \alpha_{21}\ket{interest} +\beta_{21}\ket{\widetilde{interest}}
\end{equation} 
with $\alpha_{21} = \sqrt{\frac{\lambda_2 - min(\lambda)}{max(\lambda) - min(\lambda)}}$ . 
As done with the interest dimension, we can represent each document in a Hilbert space in all the different basis corresponding to different dimensions of relevance (Steps 6 - 8).

Now in the original list of documents, suppose the user SAT-clicks on document $d_x$. We have already formed the Hilbert space for each document. The coefficients of superposition for $\ket{d_x}$ vector in a basis represent the importance of document $d_x$ with respect to the relevance dimension represented by that basis. We re-capture it by taking the projection of $\ket{d_x}$ onto the relevance vector of each basis by taking square of their inner products (Step 13), for example, $|\bra{novelty}\ket{d_x}|^2$, $|\bra{reliability}\ket{d_x}|^2$, etc. Here the notation $\bra{A}\ket{B}$ is the inner product of the vectors $A$ and $B$. It denotes the probability amplitude, which is in general a complex number. The square of probability amplitude is the probability that document $B$ is relevant with respect to dimension $A$. For real valued vectors, inner product is same as the dot product. Let $\alpha_{x1}, ..., \alpha_{x7}$ be the projections obtained. Note that the values of these projections are the same normalized scores which we calculated above. If there are more than one SAT clicked documents for a query, we average over the projection scores for each dimension (Step 15).

Thus, for a given query in a search session, we have quantitatively captured the user's cognitive state. This cognitive state or the user preference for each dimension is the average relevance score for that dimension over all SAT-clicked documents of the query. These weights are used to re-rank documents for the next query in the session, as explained in the next section.

\begin{algorithm}[H]
\caption{Capturing weights given to relevance dimensions}\label{alg:algo1}
\begin{algorithmic}[1]
\Procedure{captureWeights}{$rel$, $docsALL$, $docsSAT$} 
\ForAll{r in rel} \Comment{rel - list of 7 dimensions} 
\State $features[r] \gets getFeatures(docsALL, r)$ \Comment{Extract features from all retrieved docs for a given query}
\State $scores[r][d] \gets reRank(docsALL, features[r])$ \Comment{re-rank based on each dim and get score}
\State $normScores[r][d] \gets normalizeScores(scores[r][d])$
\ForAll{d in docsALL}
\State $\alpha[d][r] \gets normScores[d][r]$ \Comment{construct vectors}
\State $\beta[d][r] \gets 1 - |\alpha[d][r]|^2$
\EndFor
\EndFor
\ForAll{r in rel}
\State $totalWeight \gets 0$
\State $avgWeight[r] \gets 0$ 
\ForAll{d in docsSAT}
\State $w_{dr} \gets |\bra{r}\ket{d}|^2$\Comment{Take projections($|\alpha[d][r]|^2$)}
\State $totalWeight \gets totalWeight + w_{dr}$
\EndFor
\State $avgWeight[r] \gets totalWeight/|docsSAT|$ \Comment{Only SAT clicks considered}
\EndFor 
\State \textbf{return} $avgWeight$  \Comment{User's importance to each dimension}
\EndProcedure
\end{algorithmic}
\end{algorithm}

\section{Experiment and analysis}
We use the same datasets as used in \cite{ASI:Jingfei}. The first one is the query logs of the Bing search engine and the second one is the combined session tracks of TREC 2013 and TREC 2014. While the Bing query logs contain information about each query in a session, the TREC dataset only contains the data about the last query for each session. The relevance criteria for Bing logs is SAT clicks and for TREC data we consider relevance grades of 1 and above to correspond to relevant documents. In Section 3, we captured the user's dimensional preference for a query in the form of weights. We now use these weights for the next query in the session, to take a weighted combination of the relevance scores of all seven dimensions for each document of the next query. Thus, for the new query, a new relevance score for each document is created based on the weighted dimensional preference for the previous query. We re-rank the documents according to these new scores and perform evaluation. We use the NDCG metric for evaluation and compare the values with those obtained in \cite{ASI:Jingfei}.

We also performed an initial analysis of the data to support our hypothesis that some combination of relevance dimensions are preferred by the user in a search session. For some randomly sampled 4837 sessions of the Bing query logs, we found that in 3910 or 80.84 percent of the sessions, one of the top three dimensions for the first query of the session  remains in the top three for all the queries of the session. Figure 2.a is the snapshot of one such session showing that the "Reliability" remains the top dimension throughout. Figure 2.b shows 20 consecutive sessions for TREC data.

\begin{figure*}[t]
    \begin{subfigure}[t]{0.35\textwidth}
        \includegraphics[width=\textwidth]{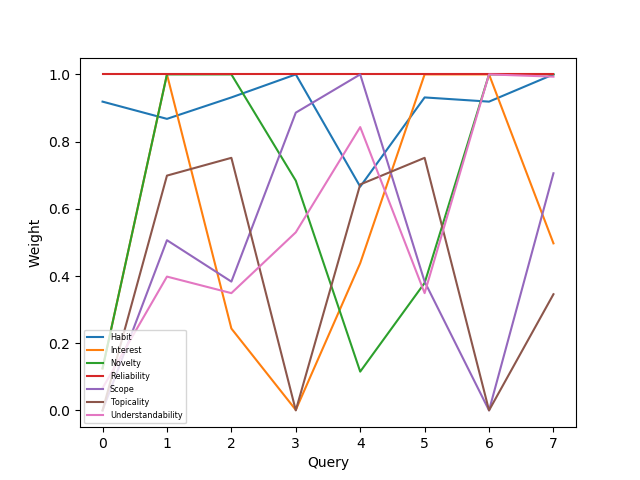}
        \caption{Figure 2.a}
    \end{subfigure}
    \begin{subfigure}[t]{0.35\textwidth}
        \includegraphics[width=\textwidth]{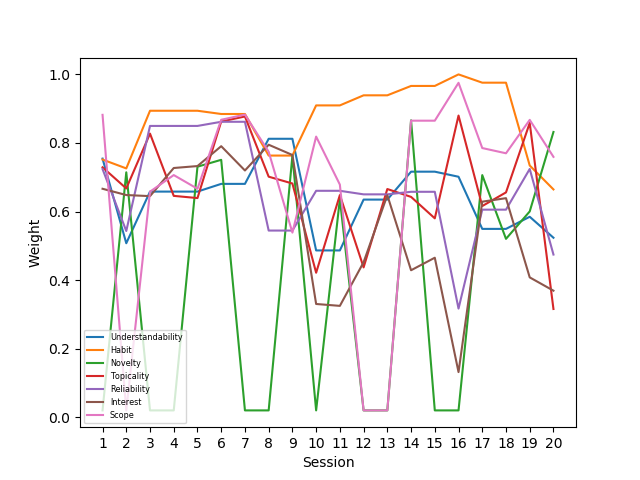}
        \caption{Figure 2.b}
    \end{subfigure}
\end{figure*}

\section{Results and discussion}
We summarize the evaluation results for Bing query logs in Table 1. In the paper \cite{ASI:Jingfei}, re-ranking by using the features of "Reliability" dimension gives the best performance for Bing data. However we show that the results obtained by using the weighted combination gives slightly better results (Table 1). The improvement is not significant but the fact is that weighted combination is a general method which will work for other datasets too. Similar to the results reported in \cite{ASI:Jingfei}, for TREC data, it is not "Reliability" but "Interest" which comes out as the best dimension for re-ranking (Table 2). Therefore one cannot use a fixed relevance dimension for ranking the documents. Table 2 shows that our weighted combination approach also gives improved performance for the TREC data. It is to be noted that TREC data contains information about the last query of each session, and not all the queries. Thus our weighted approach uses the captured weights of the last query of a session to re-rank the documents for the last query of the next session. Improvement over the best result (corresponding to Interest) means that the weighted combination method for ranking works across sessions as well. This indicates that dimensional preference is not only dependent upon the task, but user might have an intrinsic preference for some dimensions as well. Also note that the "Topicality" scores correspond to a traditional baseline model as we use tf-idf and BM25 as features for the "Topicality" dimension. 

\begin{table}[h!]
\begin{center}
\resizebox{\columnwidth}{!}{%
 \begin{tabular}{||c|c|c|c|c||} 
 \hline
 Dimension & NDCG@1 & NDCG@5 & NDCG@10 & NDCG@ALL \\ [0.5ex] 
 \hline\hline
 Habit & 0.3772 & 0.5958 & 0.6533 & 0.6645 \\ 
 \hline
 Interest & 0.4574 & 0.6178 & 0.6844 & 0.6955  \\
 \hline
 Novelty & 0.4110 & 0.6025 & 0.6688 & 0.6783 \\
 \hline
 \textbf{Reliability} & \textbf{0.6457} & \textbf{0.7687} & \textbf{0.8038} & \textbf{0.8110} \\
 \hline
 Scope & 0.2501 & 0.4692 & 0.56156 & 0.5726 \\ 
 \hline
 Topicality  & 0.2001 & 0.4486 & 0.5352 & 0.5482 \\
 \hline
 Understandability  & 0.2782 & 0.4968 & 0.5867 & 0.5971 \\
 \hline
 \textbf{Weighted Combination} & \textbf{0.6552} & \textbf{0.7814} & \textbf{0.8127} & \textbf{0.8189} \\
\hline
 \end{tabular}
 }

\end{center}
\caption{Bing logs evaluation}
\end{table}

\begin{table}[h!]
\begin{center}
\resizebox{\columnwidth}{!}{%
 \begin{tabular}{||c|c|c|c|c||} 
 \hline
 Dimension & NDCG@1 & NDCG@5 & NDCG@10 & NDCG@ALL \\ [0.5ex] 
 \hline\hline
 Habit & 0.0989 & 0.1406 & 0.1418 & 0.1592 \\ 
 \hline
 \textbf{Interest} & \textbf{0.1981} & \textbf{0.2126} & \textbf{0.2242} & \textbf{0.1831}  \\
 \hline
 Novelty & 0.0966 & 0.1180 & 0.1316 & 0.1557 \\
 \hline
 Reliability & 0.1120 & 0.1333 & 0.1431 & 0.1614 \\
 \hline
 Scope & 0.1318 & 0.1526 & 0.1647 & 0.1671 \\ 
 \hline
 Topicality  & 0.1459 & 0.1520 & 0.1887 & 0.1701 \\
 \hline
 Understandability  & 0.1653 & 0.1913 & 0.1878 & 0.1764 \\
 \hline
 \textbf{Weighted Combination} & \textbf{0.2364} & \textbf{0.2663} & \textbf{0.2729} & \textbf{0.1944} \\
\hline
 \end{tabular}
 }

\end{center}
\caption{TREC data evaluation}
\end{table}

\section{Conclusion and future work}
We have thus shown that capturing user's weights for relevance dimensions and ranking based on the combination of these weights leads to a better performance than using only one of the dimensions. The need for a Hilbert space framework is inspired by the fact that some relevance dimensions are incompatible for some documents. A document may not have high relevance weights for both "Novelty" and "Habit" dimensions at the same time. The more relevant it is in the "Novelty" dimension, the less relevant it will be in the "Habit" dimension. This is similar to the Uncertainty Principle in Quantum Theory. We therefore model each relevance dimension as a different basis. For some documents, the basis might coincide, but in general there is incompatibility between relevance dimensions which leads to interference and order effects~\cite{Busemeyer:2012:QMC:2385442, ASI:Order}. For example, a user may find a document less reliable due to its source, but when the user considers the "Topicality" dimension and reads it, it might remove the doubts about the reliability. Thus  "Topicality" interferes with "Reliability" in relevance judgment. Such order effects were investigated in \cite{10.3389/fpsyg.2014.00612} through user studies. We intend to investigate such cognitive phenomena in real world data, and the Hilbert space representation described in this paper forms a solid basis to carry out such experiments in the future. 

\begin{acks}
This work is funded by the European Union's Horizon 2020 research and innovation programme under the Marie Sklodowska-Curie grant agreement No 721321.
\end{acks}

\bibliographystyle{ACM-Reference-Format}
\bibliography{bibliography}

\end{document}